\documentclass[twocolumn,showpacs,preprintnumbers,amsmath,amssymb,aps,prl]{revtex4}

\usepackage{graphicx}
\usepackage{dcolumn}
\usepackage{bm}

\begin{document}

\title{Symmetry and Self-Organization in Complex Systems}

\author{Ben D. MacArthur}
\email{bdm@soton.ac.uk}
 \affiliation{Bone and Joint Research Group, University of Southampton, Southampton General Hospital, Southampton, SO16 6YD, UK} 
\author{James W. Anderson}
\affiliation{School of Mathematics, University of Southampton, Southampton, SO17 1BJ, UK}

\date{\today}

\begin{abstract}
We show that, in contrast to classical random graph models, many real-world complex systems -- including a variety of biological regulatory networks and technological networks such as the internet -- spontaneously self-organize to a richly symmetric state. We consider the organizational origins of symmetry and find that growth with preferential attachment confers symmetry in highly branched networks. We deconstruct the automorphism group of some real-world networks and find that some, but not all, real-world symmetry can be accounted for by branching. We also uncover an intriguing correspondence between the size of the automorphism group of growing random trees and the random Fibonacci sequences.  
\end{abstract}


\maketitle

\section{\label{sec:intro} Introduction}
The use of complex networks to model the underlying topology of real-world complex systems has attracted much research interest\cite{newman, albertSM}. Formally, the study of complex networks is underpinned by the mathematical theory of graphs\cite{newman}, and it is a classical result of random graph theory that almost all large random (Erd{\"o}s-R{\'e}nyi) graphs are asymmetric\cite{bollobasRG}. In this letter we consider the symmetry structure of a variety of real-world networks and find that, in contrast to the classical random graph models, a rich degree of symmetry is ubiquitous in real-world complex systems.

Symmetry is a source of great abstract beauty\cite{weyl} and is of great practical use in simplifying complex problems\cite{golubitsky}. Accordingly, symmetry is a well-established cornerstone of many branches of physics\cite{wigner, zee}. However, despite a well developed abstract theory of graph symmetry\cite{cameron, lauri, dicks}, the study of symmetry in real-world complex networks has been limited. 

In fact, when present, symmetry provides a powerful tool to understand network structure and function. For example, symmetry can strongly affect a network's eigenvalue spectrum\cite{cvetkovic, chung} -- by giving rise to degenerate eigenvalues, for instance\cite{lauri} -- causing it to deviate from those of the classical ensembles of random matrix theory\cite{farkas, mehta}. Furthermore, symmetry can also have a profound influence on the dynamics of processes taking place on networks\cite{golubitsky, palacios} and may be exploited to reduce the computational complexity of network algorithms\cite{grape,kocay}. 

Mathematically, a network is a graph, $G=G(V,E)$, with vertex set, $V$ (of size $N$), and edge set, $E$ (of size $M$) where $2$ vertices are connected if there is an edge between them. Here we consider the symmetry structure of various real-world networks via their automorphism group. An automorphism is a permutation of the vertices of the network which preserves adjacency. The set of automorphisms under composition forms a group, $\Gamma_G$, of size $a_G$\cite{bollobasMGT}. If the network is a multi-digraph, we remove weights and directions and consider the automorphism group of the \textit{underlying} graph. A network is said to be symmetric (respectively asymmetric) if its underlying graph has a nontrivial (respectively trivial) automorphism group and the degree of network symmetry is quantified by $a_G$. In order to compare networks of different sizes with each other we also consider the quantity $r_G= (a_G/N!)^{1/N}$ which measures symmetry relative to maximum possible symmetry (the complete graph on $N$ vertices and its complement, the empty graph on $N$ vertices, are the most symmetric graphs, both having $a_G=N!$). Other measures of graph symmetry are considered elsewhere\cite{lauri,dekker}. Here, the \texttt{nauty} program\cite{mckay} -- which includes one of the most efficient graph isomorphism algorithms\cite{foggia} -- is used to calculate the size and structure of the various automorphism groups. 

Table \ref{auttable} gives the size of the automorphism group of some real-world complex networks, all of which are highly symmetric. Since classical Erd{\"o}s-R{\'e}nyi random graphs are generally asymmetric\cite{bollobasRG}, this rich degree of symmetry is surprising and begs an explanation. The ubiquity of symmetry in disparate real-world systems suggests that it may be related to generic self-organizational principles. In order to begin to investigate the relationship between symmetry and self-organization we consider how the processes of growth and preferential attachment affect system symmetry. 

\begin{table*}
\begin{center}
\begin{ruledtabular}
\begin{tabular}[c]{l  c  c  c  c}
Network & N & M & $a_G$ & $r_G$\\
\hline 
Human B Cell Genetic Interactions\cite{basso} & $5,930$ & $64,645$ & $5.9374 \times 10^{13}$ & $4.6044 \times 10^{-4}$\\
\textit{C. elegans} Genetic Interactions\cite{zhong} & $2,060$ & $18,000$ & $6.9985 \times 10^{161}$ & $1.5776 \times 10^{-3}$ \\
BioGRID datasets\cite{biogrid}: \\
\hspace{3.2cm} Human & $7,013$ & $20,587$ & $1.2607 \times 10^{485}$ & $4.5418\times 10^{-4}$\\
\hspace{3.2cm} \textit{S. cerevisiae} & $5,295$ & $50,723$ & $6.8622 \times 10^{64}$ & $5.2753 \times 10^{-4}$\\
\hspace{3.2cm} \textit{Drosophila} & $7,371$ & $25,043$ & $3.0687 \times 10^{493}$ & $4.2993\times 10^{-4}$\\  
\hspace{3.2cm} \textit{Mus musculus} & $209$ & $393$ & $5.3481 \times 10^{125}$ & $5.1081 \times 10^{-2}$\\
Internet at the Autonomous Systems Level\cite{caida} & $22,332$ & $45,392$ & $1.2822 \times 10^{11,298}$ & $3.9009\times 10^{-4}$\\
US Power Grid\cite{wattsSW} & $4,941$ & $6,594$ & $5.1851 \times 10^{152}$ & $5.9011\times 10^{-4}$\\
\end{tabular}
\end{ruledtabular}
\end{center}
\caption{\label{auttable}\small{\textbf{The size of the automorphism group of some real-world networks.} The size of the automorphism group of the giant component is given (to $5$ significant figures) which, in all cases, contains at least $93\%$ of the vertices in the network. Giant components were extracted using \texttt{Pajek}\cite{pajek}.}} 
\end{table*} 

\section{\label{sec:self} Symmetry and Self-Organization}
Based upon the observation that many real-world networks are continuously growing and that new vertices often show a preference for attachment to more highly connected vertices, Barab{\'a}si and Albert proposed a simple model which accounted for the origin of the power-law vertex degree distribution often seen in many real-world networks\cite{barabasiPA}. Power-law degree distributions have been widely discussed\cite{li} and a number of variations of the Barab{\'a}si-Albert model have been suggested\cite{bollobasLCD2,pennock}. Here, we consider a simple variation based upon that proposed by Pennock \textit{et al}\cite{pennock}. We start with the complete graph on $k$ vertices as an initial seed. At each time step the system is updated by introducing a single new vertex and $k$ new edges which connect the new vertex to those already in the system without allowing multiple edges. The probability that vertex $v_i$ is chosen at time $t$ to attach the new vertex to is given by
\begin{equation} \label{attachprob}
P(v_i,t) = \alpha \frac{d_i(t)}{2M(t)} + (1-\alpha)\frac{1}{N(t)}
\end{equation} 
where $M(t) = k(k-1)/2+kt$ is the total number of edges in the system at time $t$, $d_i(t)$ is the degree of vertex $v_i$ at time $t$, $N(t)=k+t$ is the total number of vertices in the system at time $t$ and $0\leq \alpha \leq 1$. Thus, when $\alpha = 0$ new vertices are attached to old in a purely uniformly random way, and when $\alpha = 1$ new vertices are attached to old in a purely preferential way. 

\begin{figure}
\includegraphics[width=0.23\textwidth]{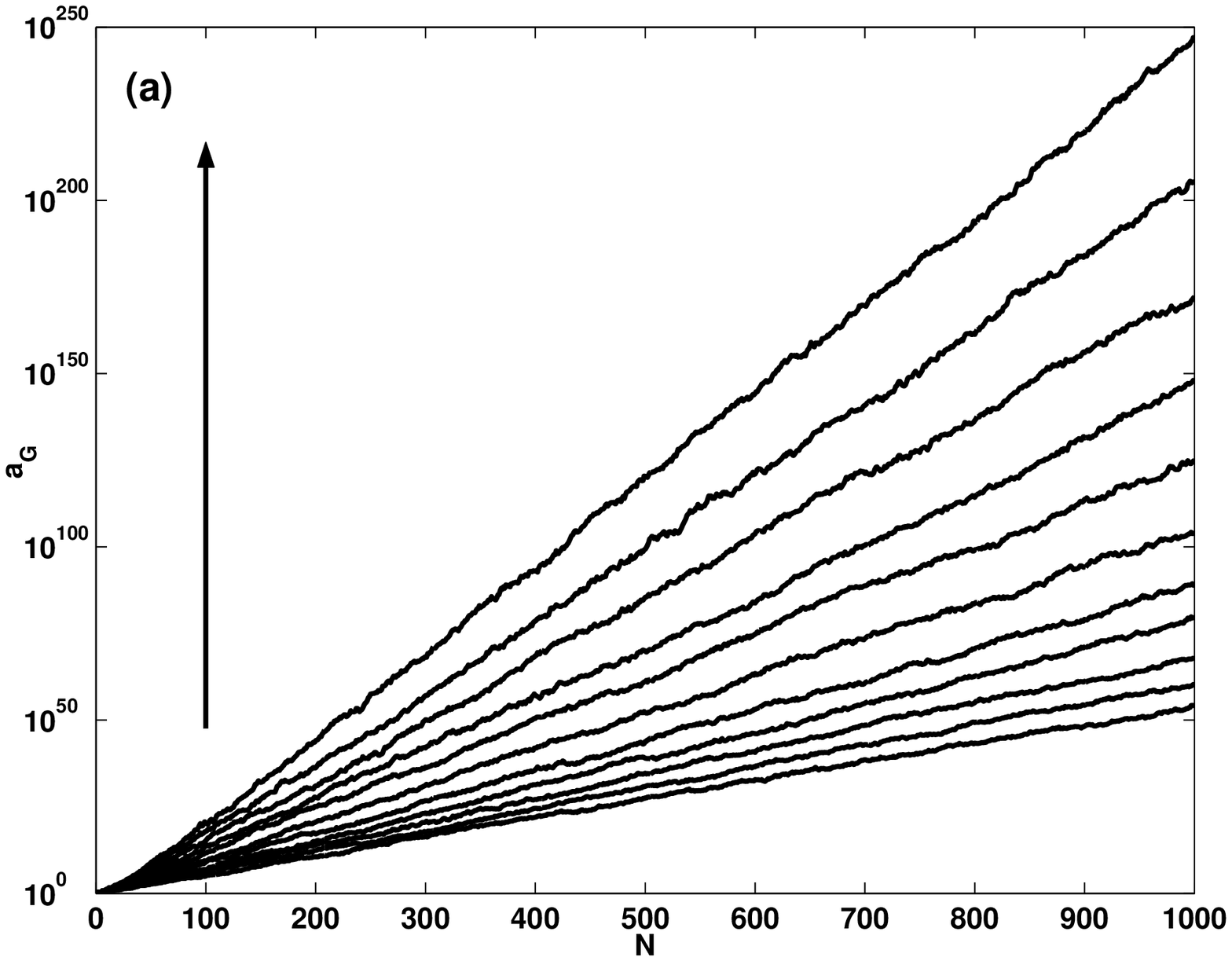}
\includegraphics[width=0.23\textwidth]{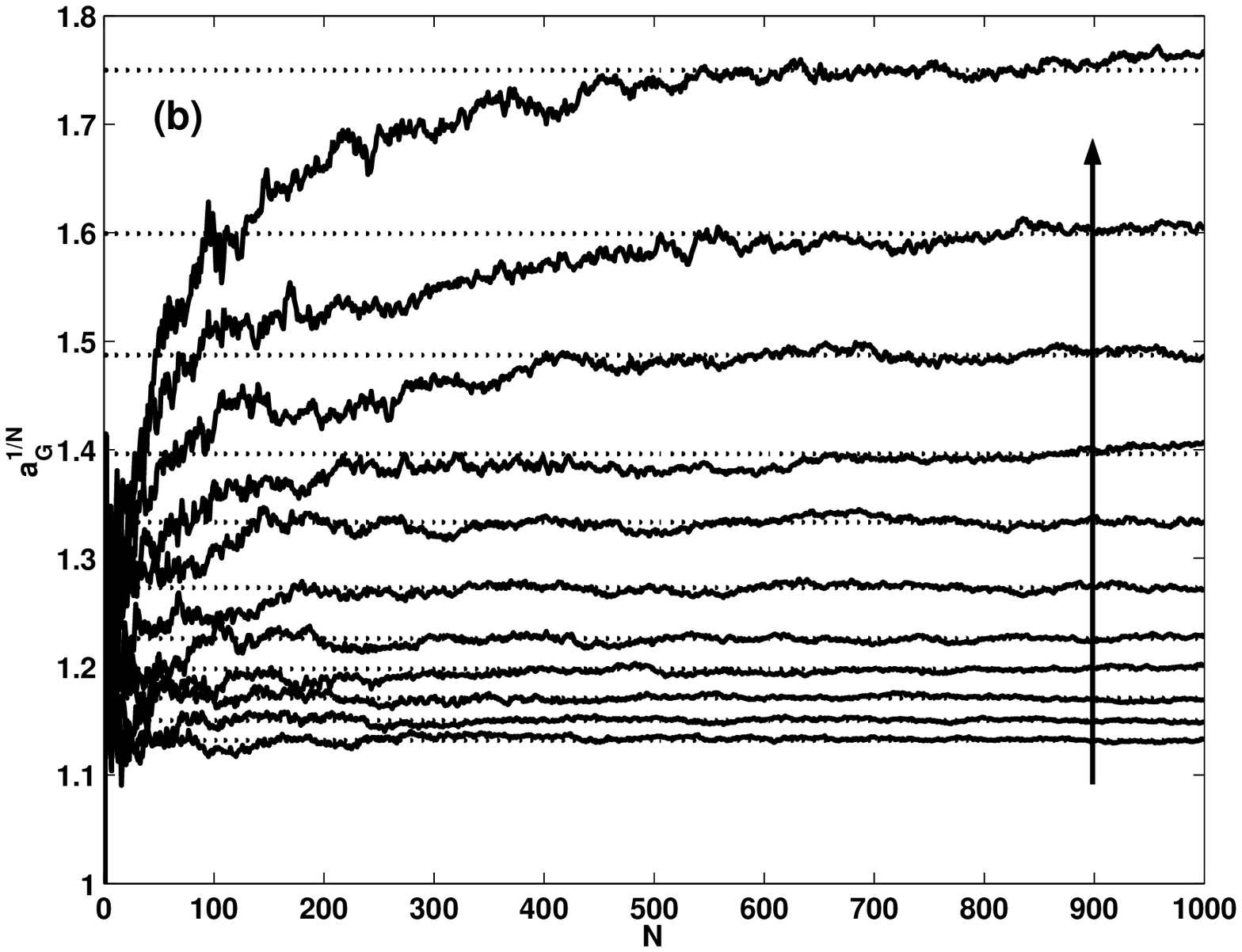}
\includegraphics[width=0.22\textwidth]{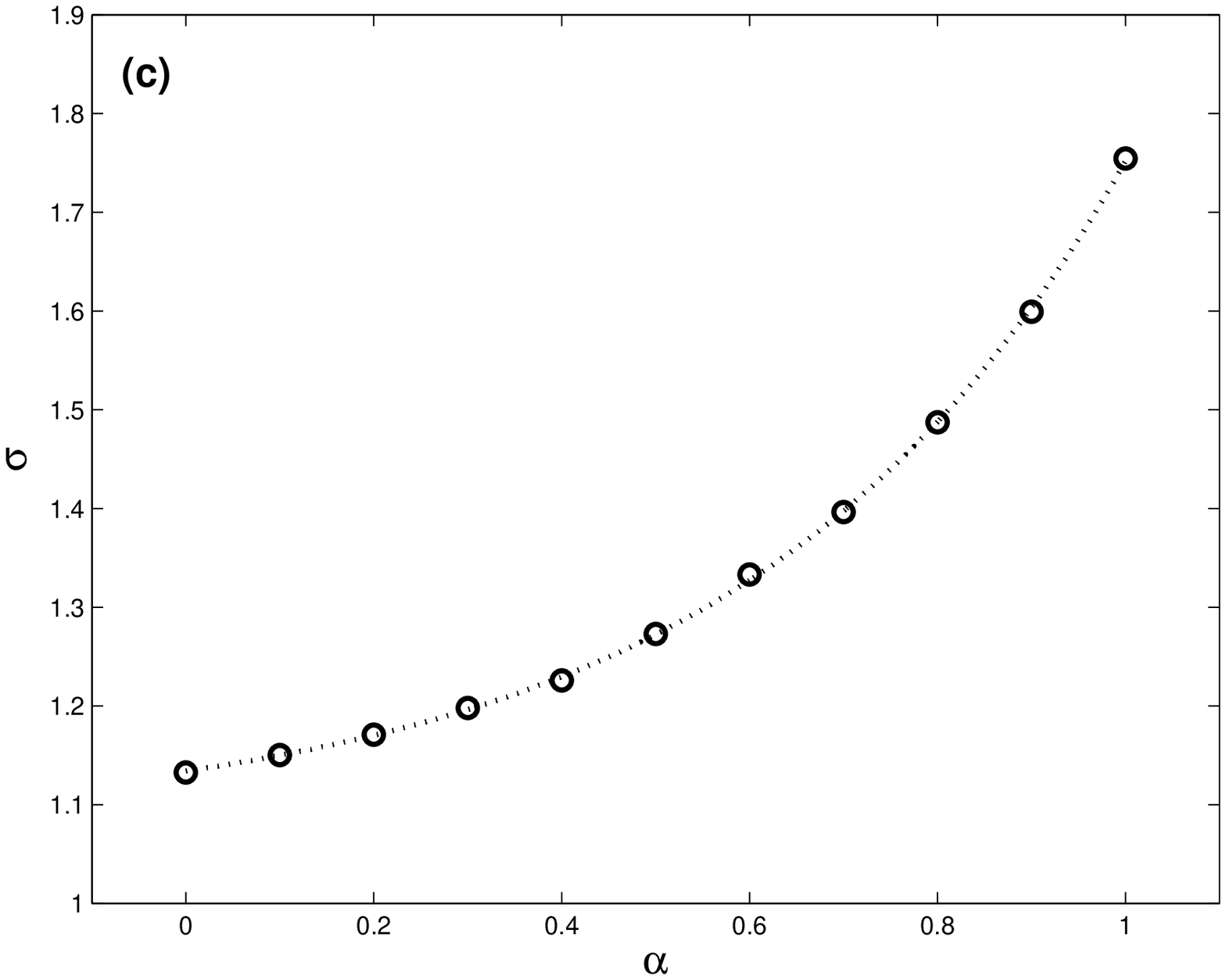}
\includegraphics[width=0.23\textwidth]{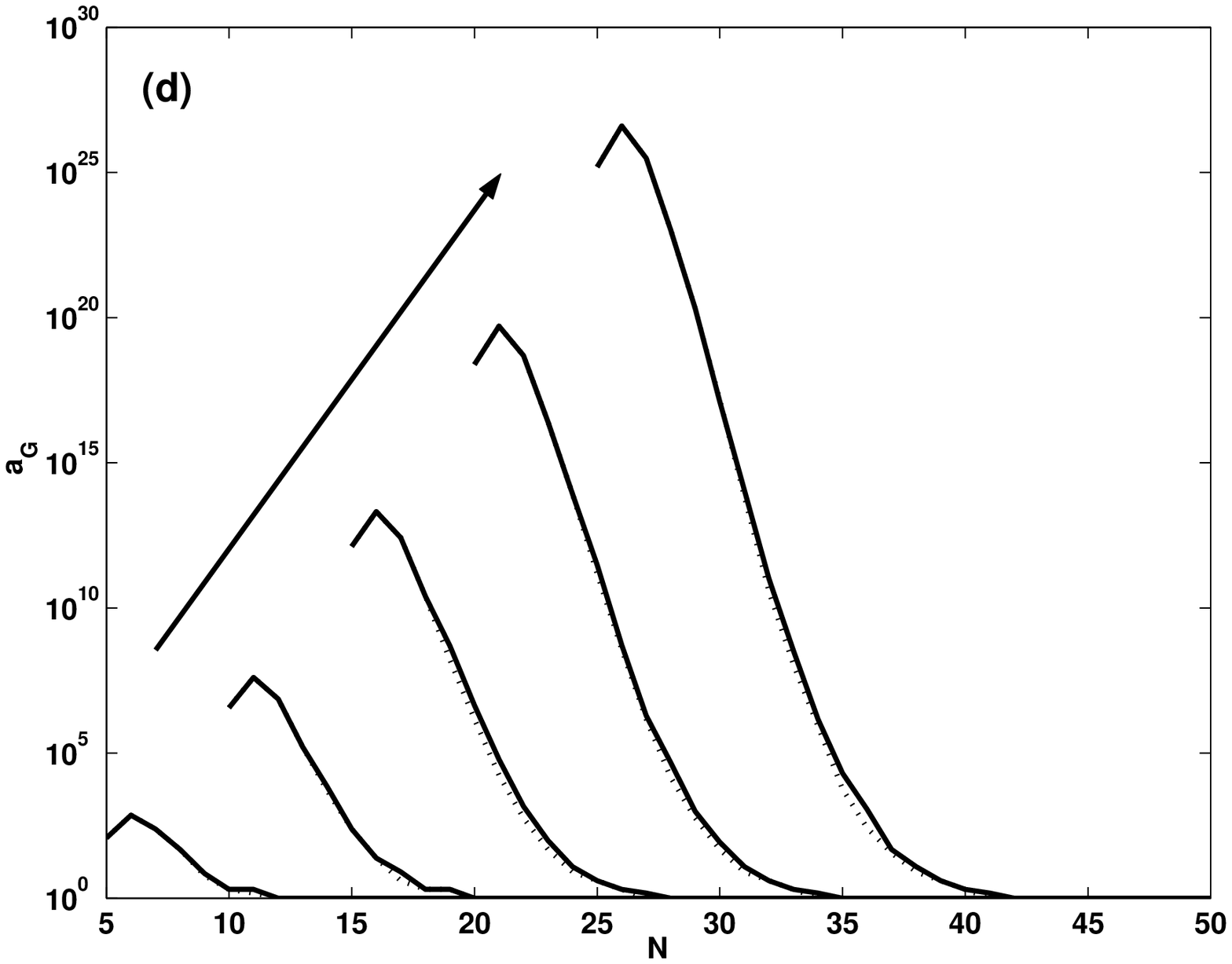}
\caption{\label{prefattach}\small{\textbf{The effect of preferential attachment on network symmetry.} (a) Growth of the automorphism group of growing random trees (k=1). Median trajectories over 50 Monte-Carlo simulations for each $\alpha$ from $0-1$ in increments of $0.1$ are shown (arrow shows increasing $\alpha$). For all $\alpha$ the growth of the automorphism group shows a clear exponential trend. (b) When $k=1$, $\lim_{N \to \infty} a_G^{1/N} =  \sigma(\alpha)$ for all $\alpha$ (median trajectories over $50$ simulations are shown). The top dotted line is $y=(1.132+0.618)^x$ and the bottom dotted line is $y=1.132^x$. (c) When $k=1$, $\sigma(\alpha)$ grows exponentially with $\alpha$ showing that, for `tree-like' growth, preferential attachment increases system symmetry. (d) Growth of the automorphism group for $k \geq 2$. Median trajectories over 50 simulations for $k=5,10,15,20,25$ are shown (arrow gives increasing $k$). Full lines give trajectories for $\alpha=1$, dotted lines give trajectories for $\alpha=0$. When $k \geq 2$ the networks quickly become asymmetric, with preferential attachment having little effect on system symmetry.}}
\end{figure}

Two distinct patterns of behavior emerge dependent upon $k$, the number of new edges associated with each new vertex. When $k=1$, $a_G$ grows exponentially and the system becomes increasingly symmetric (see Fig.\ref{prefattach}a); while for $k\geq 2$, $a_G$ rapidly falls to zero, and the system becomes asymmetric for large time (see Fig.\ref{prefattach}d). This discrepancy results since the two cases ($k=1$ and $k\geq2$) represent two fundamentally different modes of growth. When $k=1$ every new vertex is attached to the network with only one new edge so cycles cannot arise and the network is always a tree. Alternatively, when $k \geq 2$ new cycles are generated with each new vertex addition, and the network is never a tree. 

\begin{figure}
\includegraphics[width=0.23\textwidth]{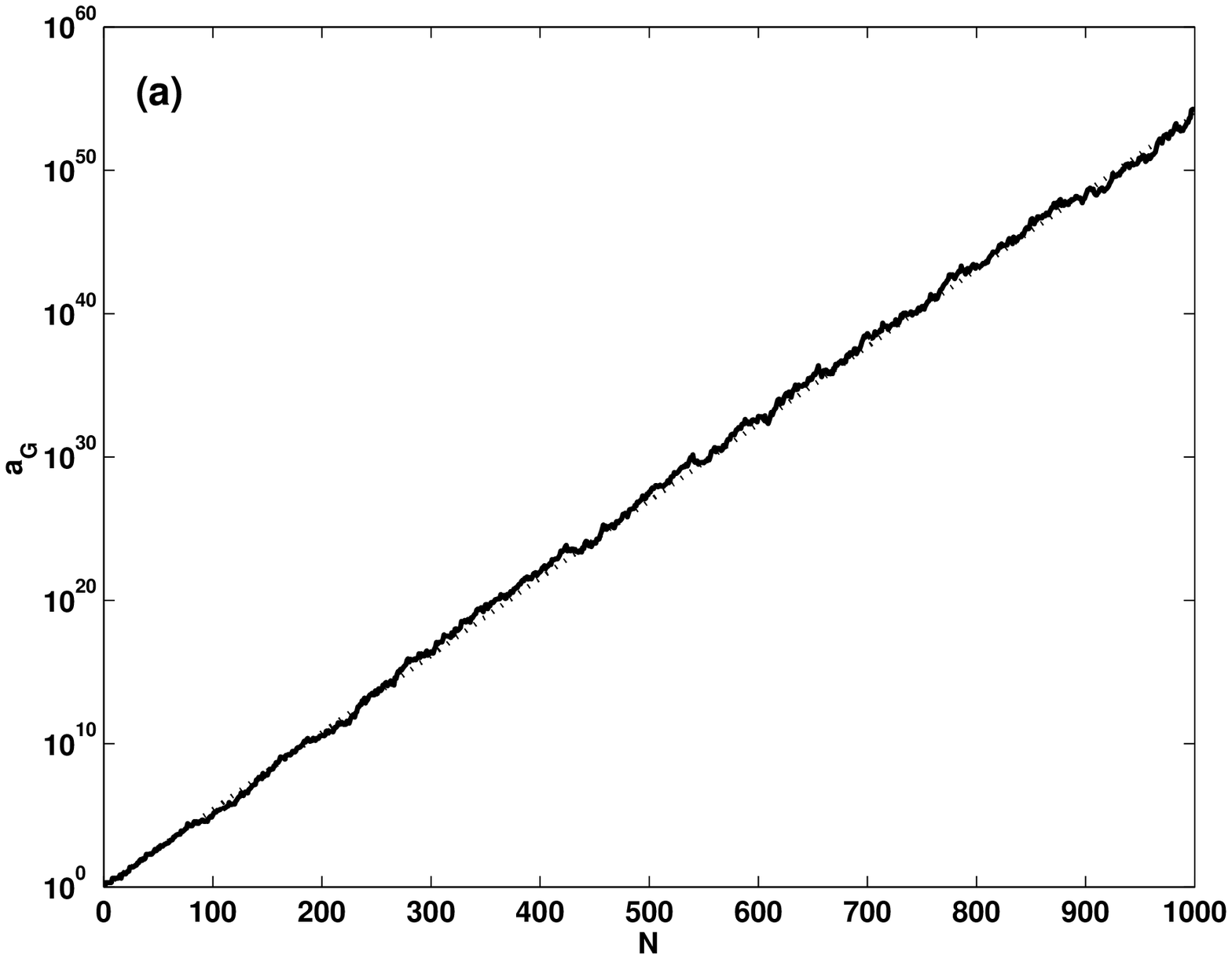}
\includegraphics[width=0.23\textwidth]{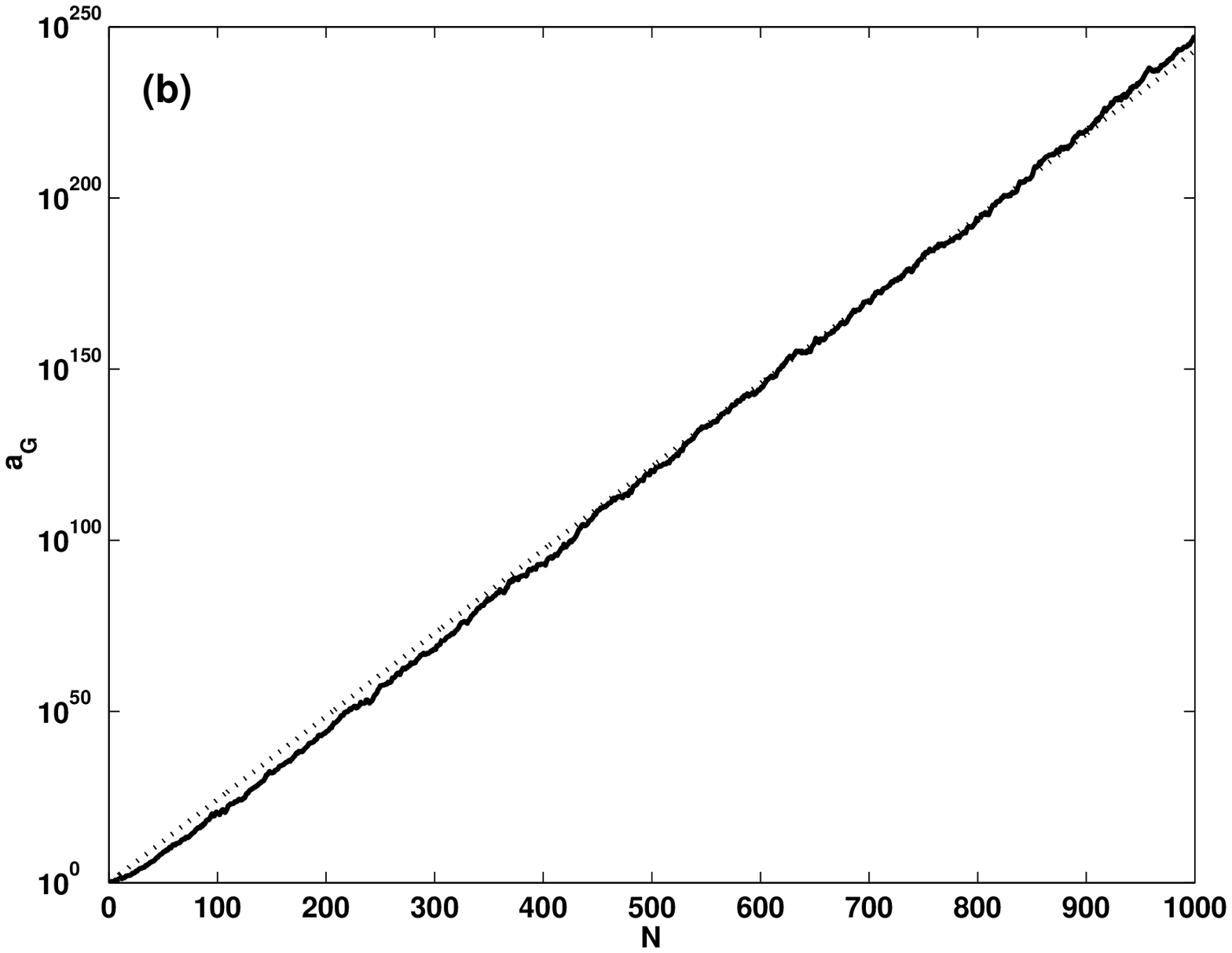}
\caption{\label{randfib}\small{\textbf{Growth of tree automorphism groups and the random Fibonacci sequences}. In both panels, $k=1$. (a) The median trajectory for $\alpha = 0$ (full line) averaged over $50$ simulations and the line $y=(1.132)^x$ (dotted). (b) The median trajectory for $\alpha =1$ (full line) averaged over $50$ simulations and the line $y=(1.132+0.618)^x$ (dotted).}}
\end{figure}

Parenthetically, these observations also suggest an intriguing relationship between the size of the automorphism group in growing random trees and the random Fibonacci sequences. The familiar Fibonacci sequence is generated by taking $f_1 =1$, $f_2=1$ and $f_i = f_{i-1} + f_{i-2}$ for $i\geq3$. Similarly, the random Fibonacci sequences are generated by taking $r_1 =1$, $r_2=1$ and $r_i = \pm r_{i-1} \pm r_{i-2}$ for all $i\geq3$ where each $\pm$ is chosen independently with probability $0.5$. Viswanath showed that $\lim_{i \to \infty} \vert r_i \vert^{1/i} =  v = 1.13198824 \ldots$\cite{viswanath}. Interestingly, our numerical simulations suggest that when $k=1$, $\lim_{N \to \infty} a_G^{1/N} = \sigma(\alpha)$ for all $\alpha$ where $\sigma(\alpha)$ is a constant (the Lyapunov constant\cite{embree}) with $\sigma(0) = v$ and $\sigma(1)= v + 1/\phi$, where $\phi = 1/2(1+\sqrt{5})$ is the golden ratio (see Fig. \ref{prefattach}b and Fig. \ref{randfib}). 

Although the majority of large random Erd{\"o}s-R{\'e}nyi graphs are asymmetric\cite{bollobasRG}, it is common for large random trees to exhibit a high degree of symmetry\cite{harary}. Intuitively, this is because the absence of cycles in trees means that the number of tree configurations are combinatorially restricted (the set of trees on $N$ vertices is a thin subset of the set of graphs on $N$ vertices) and this restriction can force repetition of identical branches from the same fork, endowing the tree with symmetries. For example, every tree contains at least two vertices of degree $1$\cite{hararyGT}. These vertices are called leaves and are the end points of branches. Whenever a branch ends in $2$ or more leaves an automorphism naturally arises by permutation of these leaves while holding all other vertices in the tree fixed. 

Consequently, we conclude that the rich degree of symmetry seen in the $k=1$ case is due to the fact that these networks grow as trees, rather than the mode of attachment of new vertices \textit{per se}. However, we find that, given that the network is growing as a tree, preferential attachment increases network symmetry (see Fig.\ref{prefattach}a-c). A heuristic explanation for this is that preferential attachment introduces a bias toward multiple short branches and, probabilistically, short branches are more likely to be repeated about the same fork than longer branches. To see this consider the following two limiting cases: (1) in which new vertices are always attached to the \textit{most} highly connected vertex. Starting from a single vertex this gives a $t$-star, $S_t$, at time $t$ where one vertex has degree $t$, all others have degree $1$ and there are $t$ branches of length $1$. In this case, $a_{S_t}=(t-1)!$ for $t\geq3$ and the tree is almost maximally symmetric. (2) in which new vertices are always attached to the \textit{least} highly connected vertex. Starting with a single vertex this gives a path (that is, a single long branch), $P_t$, at time $t$ with $2$ vertices of degree $1$ and $t-2$ vertices of degree $2$. In this case, $a_{P_t}=2$ for all $t > 1$ and the graph is minimally symmetric. 

\begin{figure}
\includegraphics[width=0.23\textwidth]{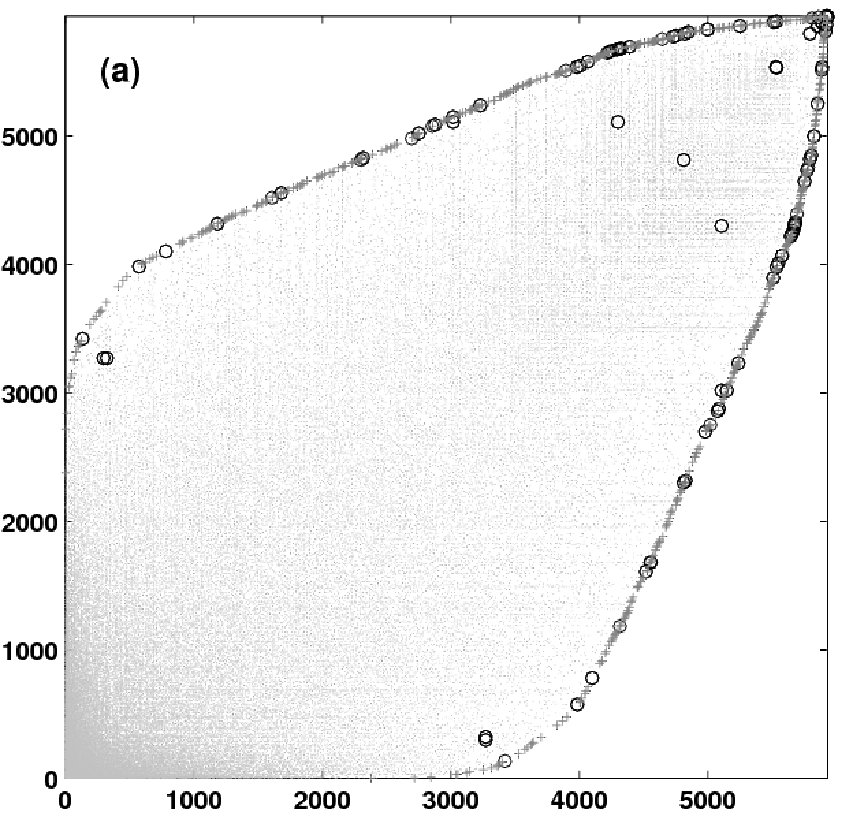}
\includegraphics[width=0.235\textwidth]{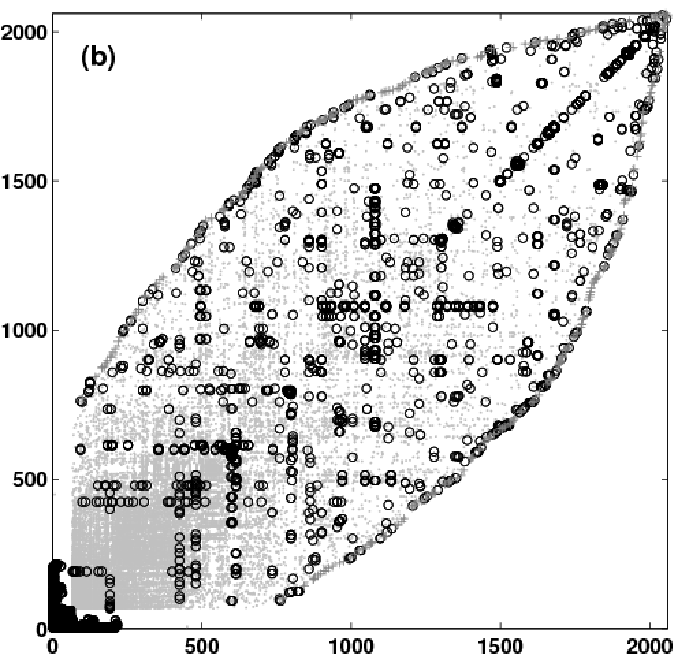}
\caption{\label{spy}\small{\textbf{Adjacency matrices of genetic interaction networks} sorted by eigenvector centrality. Edges associated with fixed leaves are dark gray crosses, those associated with non-trivial vertices are black circles and those associated with non-leaf fixed vertices are light gray points. Edges linking fixed to non-trivial vertices are colored as non-trivial. (a) The Human B cell genetic network. (b) The \textit{c. elegans} genetic network.}}
\end{figure}

These results suggest that branching may be a source of symmetry in real-world systems. In order to see if this is the case we considered the orbit structure of some real-world networks. For each vertex $v$, in a graph $G$, the \textit{orbit} of $v$ under the action of $\Gamma_G$ is the set $\Gamma_G(v) = \{\gamma v\ \in G : \gamma \in \Gamma_G \}$\cite{cameron}. When $\vert \Gamma_G(v) \vert =1$ the vertex is \textit{fixed} (or trivial) otherwise it is non-trivial. Since orbits are disjoint\cite{cameron} the symmetry structure of a network can be investigated by visualizing its adjacency matrix and applying different colors to different orbits. As illustrative examples, Fig. \ref{spy} gives the adjacency matrices of the human B cell and \textit{c. elegans} genetic regulatory networks. Here, in order further clarify network structure, the rows and columns of the adjacency matrices are also sorted in descending order by eigenvector centrality (eigenvector centrality is a measure of vertex importance in a network, and is the basis of numerous network algorithms\cite{newman}). Fig. \ref{spy}a shows that the majority of automorphisms of the B cell network are permutations of leaves indicating that, as in random trees, branching is the dominant source of symmetry in this network. By contrast, Fig. \ref{spy}b shows that the symmetry structure of the \textit{c. elegans} network is much more intricate, with symmetry arising both from branching (as evidenced by movement of the leaves) and elsewhere. For example, the $66$ most central vertices in the \textit{c. elegans} network preferentially associate with each other to form a highly cyclic, highly symmetric core. Of these $66$ vertices, half associate exclusively with other vertices in this core. Since these $33$ vertices can be permuted amongst themselves without affecting network structure, they form an orbit of length $33$ (the largest orbit in the network). The induced subgraph on these $33$ vertices is complete, and therefore has an automorphism group of size $33! = 8.6833 \times 10^{36}$ ($5$ s.f.) which constitutes a significant proportion of the symmetry of the network as a whole. The presence of this highly cyclic, highly symmetric region demonstrates that not all real-world symmetry derives from branching.   

\section{\label{sec:conclusions} Conclusions}
In summary, we have shown that many real-world networks are richly symmetric and have traced the origin of some of this symmetry to branching. Furthermore, using a simple mathematical model of network growth we have shown that preferential attachment can increase symmetry in tree-like regions by encouraging the formation of multiple short branches. However, as illustrated by the \textit{c. elegans} genetic regulatory network, not all real-world network symmetry can be accounted for by branching alone. 

Some potential benefits of graph symmetry are known (symmetry enhances tolerance to attack\cite{dekker}, for example). However, in real-world networks it is unclear whether such benefits derive from underlying organizational principles -- with symmetry being incidental -- or whether they are intrinsic to symmetry itself (growth with preferential attachment also enhances tolerance to attack\cite{albertAttack}, for example). In other words, it is unclear if symmetry is ever \textit{itself} an organizing principle in network evolution. Elucidation of the relationship between symmetry and self-organization and further investigation of the symmetry structure of real-world networks is fertile ground for future research. We anticipate that such investigations will significantly enrich our understanding of the structure and behavior of many real-world complex systems.

\end{document}